\documentclass{article}
\usepackage{spconf,amsmath,amssymb,graphicx,subcaption}

\DeclareMathOperator*{\argmin}{arg\,min}
\title{Single channel speech enhancement using outlier detection}

\name{Eunjoon Cho\thanks{This was work was done while Eunjoon Cho was at Stanford University. He is currently affiliated with Google Inc.}\footnote{sdf}$^{\star}$ \qquad Bowon Lee$^{\dagger}$ \qquad Ronald Schafer$^{\star}$ \qquad Bernard Widrow$^{\star}$}

\address{$^{\star}$ Department of Electrical Engineering, Stanford University \\
    $^{\dagger}$ Department of Electronic Engineering, Inha University} 

\begin{document}
\maketitle

\begin{abstract}
Distortion of the underlying speech is a common problem for single-channel speech enhancement algorithms, and hinders such methods from being used more extensively. A dictionary based speech enhancement method that emphasizes preserving the underlying speech is proposed. Spectral patches of clean speech are sampled and clustered to train a dictionary. Given a noisy speech spectral patch, the best matching dictionary entry is selected and used to estimate the noise power at each time-frequency bin. The noise estimation step is formulated as an outlier detection problem, where the noise at each bin is assumed present only if it is an outlier to the corresponding bin of the best matching dictionary entry.
This framework assigns higher priority in removing spectral elements that strongly deviate from a typical spoken unit stored in the trained dictionary. Even without the aid of a separate noise model, this method can achieve significant noise reduction for various non-stationary noises, while effectively preserving the underlying speech in more challenging noisy environments.
\end{abstract}

\section{Introduction} \label{sec:intro}
Single-channel speech enhancement is an underdetermined problem since only the noisy speech is available. A widely used assumption is that the speech is temporally sparse, and the noise properties vary slowly compared to speech. With this assumption, the noise power estimate at a certain frequency is updated when the spectral bin is judged to be dominated by noise. This noise estimate is used to enhance speech until another chance to update the noise estimate occurs. 
Noise estimation methods~\cite{martin01, cohen03, gerkmann12} apply different criteria on when to update the noise, and are typically used to compute the spectral gain for speech enhancement algorithms~\cite{ephraim84, scalart96}.

Given the assumption that these methods rely on, such algorithms are more susceptible to non-stationary noise. Methods that use prior knowledge of the speech and/or noise have shown to be effective under more realistic non-stationary noisy environments~\cite{srinivasan06, ellis06, ming11, duan12, sigg12}. These methods learn a representation the sources in advance, and use this to enhance the noisy speech input. Parametric models capture the sources in a compact representation by training the coefficients of a model~\cite{srinivasan06, ming11}. Non-parametric models learn representations of the sources directly from the spectrum, and are trained using vector quantization (VQ)~\cite{ellis06, christensen11}, non-negative matrix factorization (NMF) and its probabilistic counterparts~\cite{smaragdis07, duan12}, and variants of singular value decomposition (SVD)~\cite{sigg12}.

However, these methods often rely on a separate noise model~\cite{srinivasan06, sigg12, duan12, christensen11}, which limits the algorithms to work in environments that have been previously trained on. When used in new environments, the algorithm can distort the underlying speech. For offline enhancement, semi-supervised NMF~\cite{smaragdis07, mysore11} provides a solution to enhace without a separate noise model. However, for real-time speech enhancement this can be computationally expensive, since a non-stationary environment would require frequent updates of the noise bases.

Speech distortion is one of the reasons why enhancement methods are not being used more extensively in real life applications. We thus provide a framework where the speech distortion can be limited for a wide range of noisy environments, without using explicit noise models. Noise is detected when a region of the noisy speech spectrum is considered an outlier to the corresponding region of a trained speech dictionary entry. In other words, a region of the noisy speech spectrogram is detected as noise if it doesn't fit our understanding of what a typical speech spectrogram should look like. Unfortunately, without the use of a separate noise model, noise reduction for speech-shaped noise, such as babble noise, is limited. However, the benefit of the outlier framework is that it removes noise that is strongly inconsistent with the speech dictionary with higher priority. For noise that is more overlapped with speech, it focuses on preserving the underlying speech as much as possible, and thus limits the amount of speech distortion while removing less noise.

\section{Proposed method} \label{sec:method}
The overall enhancement method is a two stage process. The training stage learns a dictionary of clean speech units. The enhancement stage uses the dictionary in an outlier detection framework to reduce the noise in noisy speech inputs.

\subsection{Dictionary training} \label{sec:train}
Given a clean speech sentence, $x(n)$, the magnitude-squared Short-time Fourier Transform (STFT) is computed to estimate the power. The magnitude-squared STFT can be denoted as $|X_m(k)|^2$, where $k$ is the Discrete Fourier Transform (DFT) frequency bin and $m$ is the frame index. 

$|X_m(k)|^2$ is normalized, such that the average power of a time-frequency bin over a training sentence\footnote{The TIMIT database is used for training the dictionary, and a sentence refers to a sample utterance from the TIMIT database.} is 1. The normalized spectrum is denoted as $|\tilde X_m(k)|^2 \triangleq A\cdot|X_m(k)|^2$, where $A$ is the normalization constant.
The purpose of this stage is to compensate the amplitude difference of recordings by different speakers. This normalization, however, only corrects for a scalar multiplication of the speakers' inputs, and doesn't necessarily correct differences in perceived loudness or filtering effects due to the different recording environments. 

Patches of the normalized magnitude-squared spectra are sampled throughout the sentence. A patch can span multiple frames. For example, if a patch has a length of $L$ in the time dimension, the patch sampled at frame $m$, would consist of a sequence of magnitude-squared spectra from frame $m$ to $m+L-1$, i.e., $\{|\tilde X_m|^2, |\tilde X_{m+1}|^2, ..., |\tilde X_{m+L-1}|^2\}$.
In order to guarantee a good mix of patches in our training data, we sample patches such that there are no overlapped patches. We sample a patch beginning at every other $M$'th frame where $M > L$. 

In order to keep the input feature at a computationally reasonable size, we map the frequency bins through a sequence of triangular filter bands. The Mel or Bark scale can also be used to center the filter bands such that the lower frequencies are emphasized~\cite{cho13}. For simplicity, here we use uniformly separated filter bands. These filters are also normalized such that the power remains the same after the transformation. The output patch has a reduced dimension of $L \times N'$, where $N'$ is the number of triangular filter bands. We concatenate the end of each column in this patch, and express the patch as a column vector, $F_X$, of size $L\cdot N'$.

The sampled patches are then clustered using the \emph{k}-means algorithm. To match the dynamic range in human auditory perception,  logarithmic distance is used to cluster $F_X$. 
The cluster centroids are stored as entries in the dictionary, where each entry can be viewed as a commonly spoken speech unit. The constant $K$ determines the number of dictionary entries. 

\subsection{Speech enhancement using the outlier framework} \label{sec:enhance}

The magnitude-squared STFT is computed from the noisy speech input, $y(n)$, and patches beginning at each frame are created. A patch is passed through the same filter bands used for training, resulting in a feature vector, $F_Y$, of dimension $L \cdot N'$.

For each noisy input patch, we search for the best matching dictionary entry. Logarithmic distance is used to find the closest entry, since it was also used for clustering the training data. The cluster (or dictionary entry), $j$, that minimizes the logarithmic distance is selected as the best match.
\begin{align*}
(a^*, j^*) = \argmin_{a, j} \lVert \log F_Y - \log \left(a \cdot \bar S_j\right) \rVert^2 
\end{align*}
$\bar S_j$ is the $j$'th dictionary entry, and $a$ is a factor that corrects the amplitude difference between the speech used in training and the input noisy speech. Simply normalizing the noisy speech spectrum in advance, as we did for training, will not work since the noise will affect the normalization. The optimal $a$ for an entry $j$ can be computed as
\begin{align*}
&\frac{\partial}{\partial a} \sum_{i=1}^{L\cdot N'} \left(\log F_Y(i) - \log a - \log \bar S_j(i) \right)^2  = 0 \\
\Leftrightarrow ~~~~ & a = \exp\left(\frac{1}{L\cdot N'}\sum_{i=1}^{L\cdot N'} \log \frac{F_Y(i)}{\bar S_j(i)}\right)
\end{align*} 
The optimal $j$ is searched by iterating over the dictionary entries.
Given the best match, an estimate of the clean speech patch can be computed as $\hat F_X \triangleq a^*\bar S_{j^*}$.

If $\hat F_X$ is an accurate representation of the underlying speech, it could be used directly to replace the noisy speech. However, as shown in~\cite{ellis06}, a VQ representation of the speech spectrum is, by itself, insufficient to capture all the subtle nuances of the underlying speech. Also, without a separate noise model, distortion of the speech is likely to occur if we simply replace the noisy patch with the best dictionary entry. 

Therefore, instead of using the dictionary entries to directly quantize the noisy speech patch, we use it as a reference to estimate the noise. If the power at a noisy patch bin, $F_Y(i)$, is much greater than the corresponding bin of the best dictionary entry, $\hat F_X(i)$, it is likely that the bin is dominated by noise. The greater the deviation, the more likely it is a noise component. Instead of trying to remove all the noise, we prioritize in first removing noise components that are strong outliers to the selected dictionary entry.

\begin{figure}
\centering
\includegraphics[width=0.45\textwidth]{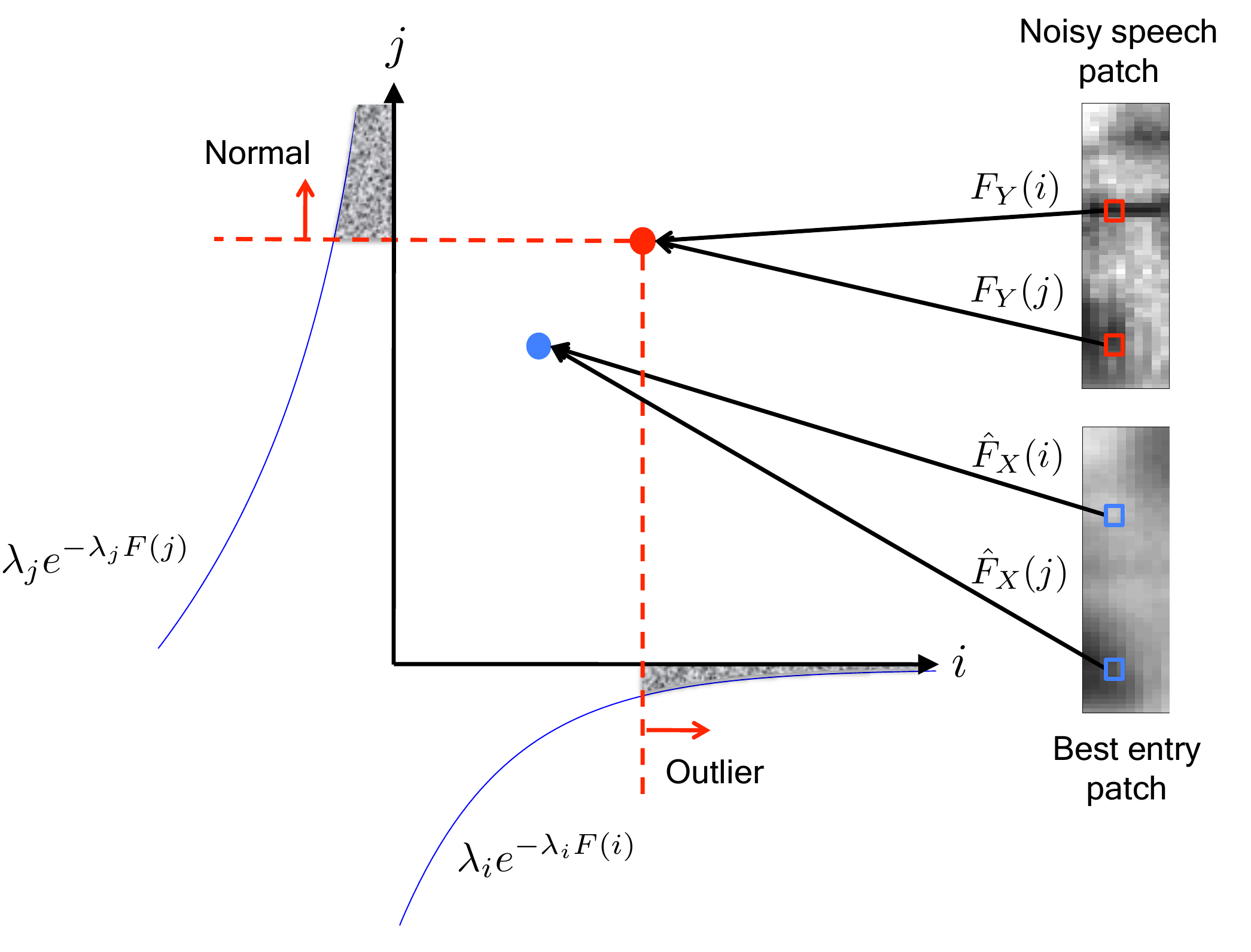}
\vspace{-1mm}
\caption{Only two dimensions ($i$, $j$) of the patches are projected here for illustration purposes. $F(i)$, $F(j)$ are exponential random variables with mean $\lambda_i \triangleq \frac{1}{\hat F_X(i)}$ and $\lambda_j \triangleq \frac{1}{\hat F_X(j)}$. Here, $F_Y(i)$ is more likely to be an outlier than $F_Y(j)$.}
\label{fig:outlier}
\vspace{-1mm}
\end{figure}

To detect whether a spectral bin, $F_Y(i)$, is an outlier, an underlying distribution for $\hat F_X(i)$ is necessary. One distribution commonly used for speech enhancement~\cite{ephraim84} is to model Fourier Transform coefficients using a complex Gaussian distribution. Under this assumption, $X_m(k)$, is complex Gaussian distributed, and $|X_m(k)|^2$ and $|\tilde X_m(k)|^2$ are exponentially distributed. We rely on computing a wideband spectrum to capture the formant envelope of the speech spectrum. With sufficient number of triangular filter bands, the power within each filter band will be approximately the same. We thus assume $F_X(i)$ is also exponentially distributed. A dictionary entry, $\bar S_j$, is the cluster centroid and each training patch, $F_X(i)$, in cluster $j$ can be viewed as independent exponential random variables with mean $\bar S_j(i)$. Since $\hat F_X(i)$ is just a scaled cluster centroid, we can compare whether $F_Y(i)$ is a good fit to this cluster by comparing it against an exponential distribution that has a mean of $\hat F_X(i)$.

Assume that $F(i)$ is a random exponential variable with mean $\hat F_X(i)$. In other words, $F(i)$ is a potential patch inside a cluster, where the cluster's centroid is $\hat F_X(i)$. The p-value of $F_Y(i)$, i.e., $P(F(i) \geq F_Y(i))$, is used to determine whether $F_Y(i)$ is an outlier or not. If this p-value is less than a threshold, $c$, it is considered an outlier. Fig.~\ref{fig:outlier} illustrates the process of evaluating whether a frequency bin of a patch is an outlier. 

If an element, $F_Y(i)$, is an outlier, we assume the noise is present and use spectral subtraction to estimate the noise. If it is not an outlier, we assume there is no noise. Specifically,
\begin{align*}
&\hat F_D(i) \\
&= \left\{ \begin{array}{ll} \max\left[F_Y(i) - \hat F_X(i),0\right] & \mbox{, if  $P\left(F(i) \geq F_Y(i)\right) < c$} \\
 0 & \mbox{, otherwise}
 \end{array} \right.
\end{align*} 
where $\hat F_D(i)$ is the estimated noise patch. The decision to ignore the case when an element is not an outlier, is to preserve the underlying speech as much as possible when in doubt. Nonetheless, the user can control the level of noise reduction by changing the threshold, $c$.

\begin{figure}
\centering
\begin{subfigure}{0.45\textwidth}
\includegraphics[width=\textwidth]{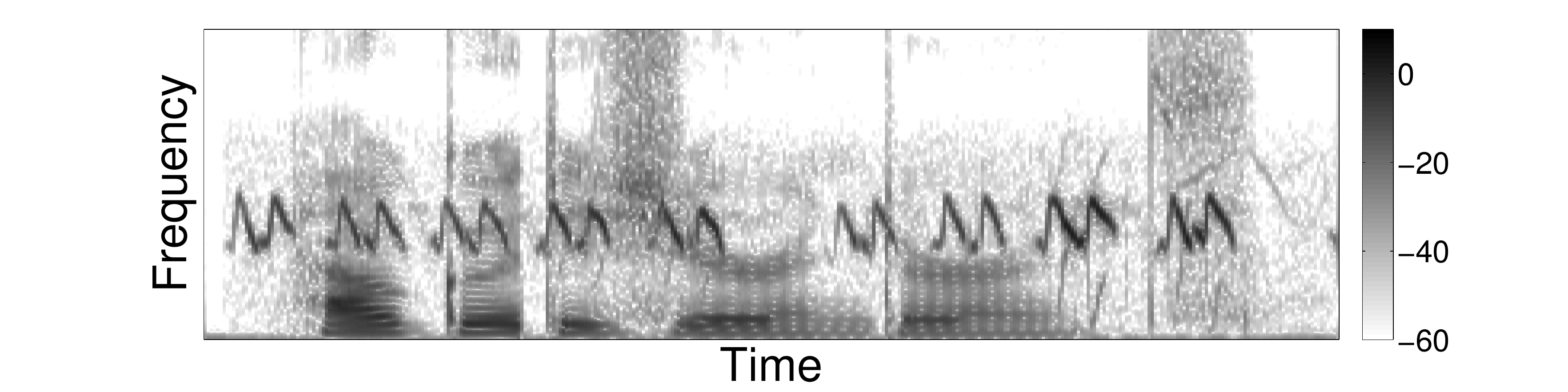}
\caption{Noisy speech}
\end{subfigure}
\begin{subfigure}{0.45\textwidth}
\includegraphics[width=\textwidth]{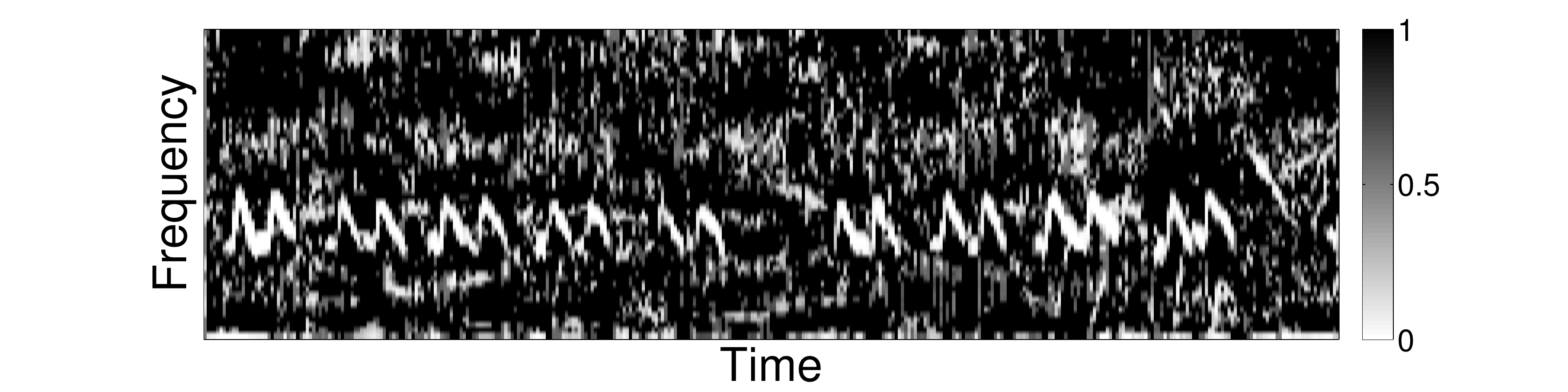}
\caption{Outlier mask for $c = 0.1$}
\end{subfigure}
\begin{subfigure}{0.45\textwidth}
\includegraphics[width=\textwidth]{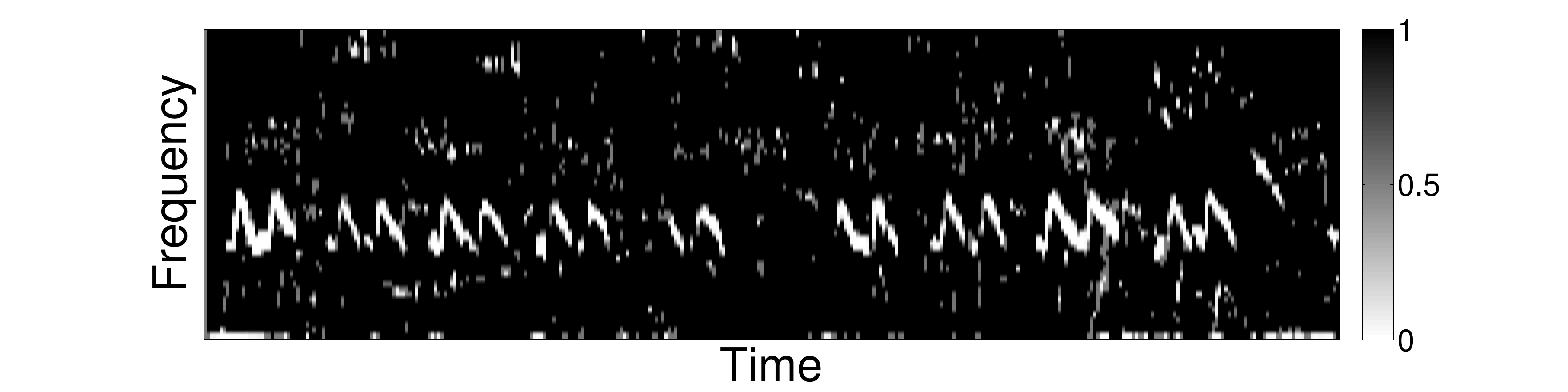}
\caption{Outlier mask for $c = 0.0001$}
\end{subfigure}
\vspace{-1mm}
\caption{Masks computed with different outlier thresholds. A larger threshold will remove more noise at the expense of more speech distortion.}
\label{fig:threshold}
\vspace{-1mm}
\end{figure}

With the estimated noise, a Wiener-like mask is computed.
\begin{align}
F_H(i) = \frac{F_Y(i) - \hat F_D(i)}{F_Y(i)}
\label{eq:mask}
\end{align}
A mask is computed for every frame, so if each patch is of length $L > 1$, these patches will overlap. For each frame, we average the $L$ different mask gains to compute the gain at that specific frame. To enhance the noisy spectrum, we need a mask in the original frequency domain. We thus transform the mask, by interpolating it with the same triangular filter bands used for analysis, so that the final mask has a frequency dimension equal to the original DFT size. This mask is then applied to the input noisy speech spectrum to get our enhanced spectrum. An example of this mask is shown in Fig.~\ref{fig:threshold}.

\section{Evaluation} \label{sec:eval}

\begin{figure*}
\centering
\begin{subfigure}{0.28\textwidth}
\includegraphics[width=\textwidth]{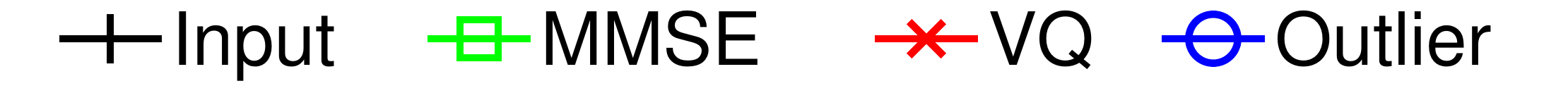}
\end{subfigure} \\
\begin{subfigure}{0.19\textwidth}
\includegraphics[height=2.9cm]{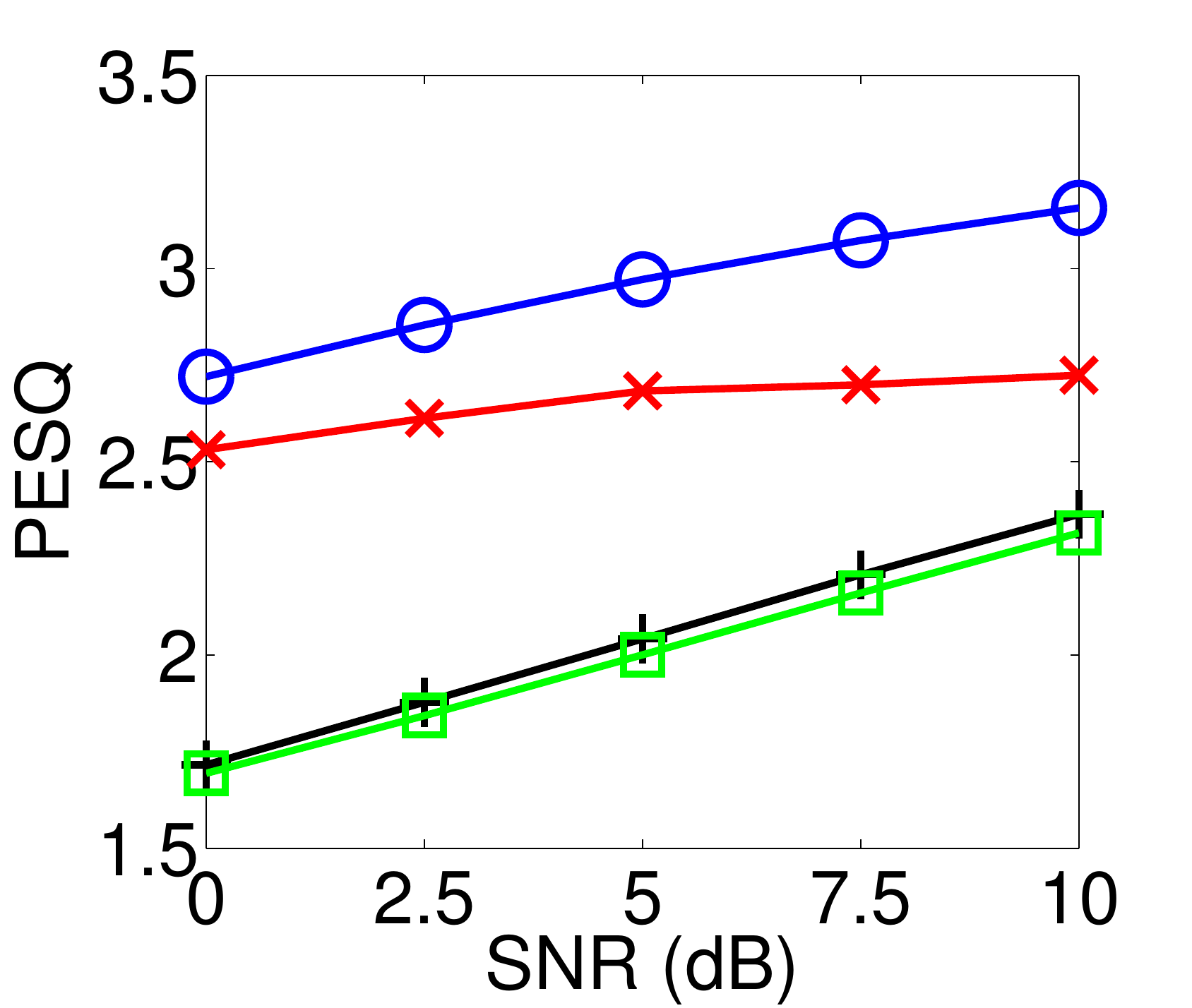}
\caption{Bird}
\end{subfigure}
\begin{subfigure}{0.19\textwidth}
\includegraphics[height=2.9cm]{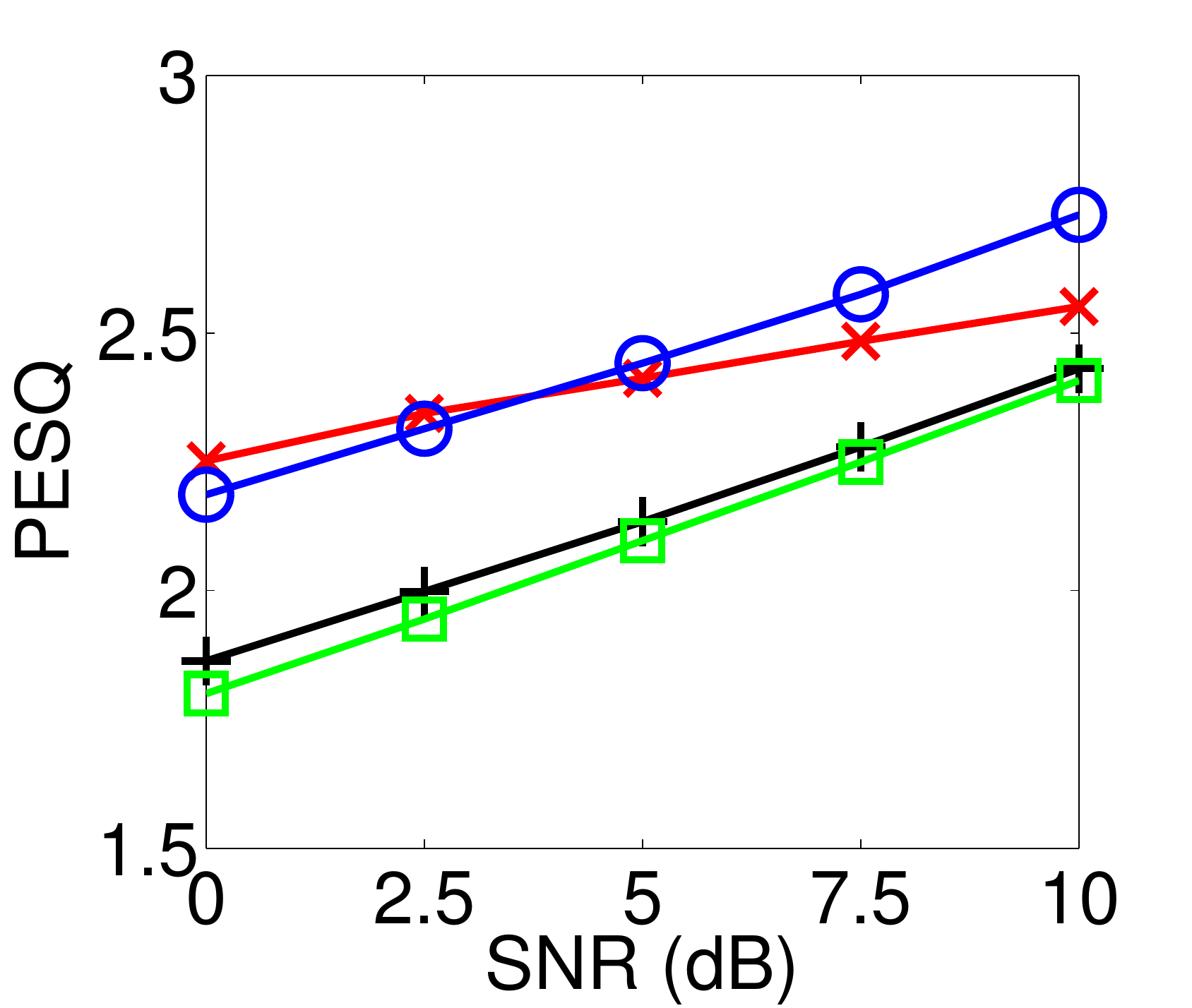}
\caption{Siren}
\end{subfigure}
\begin{subfigure}{0.19\textwidth}
\includegraphics[height=2.9cm]{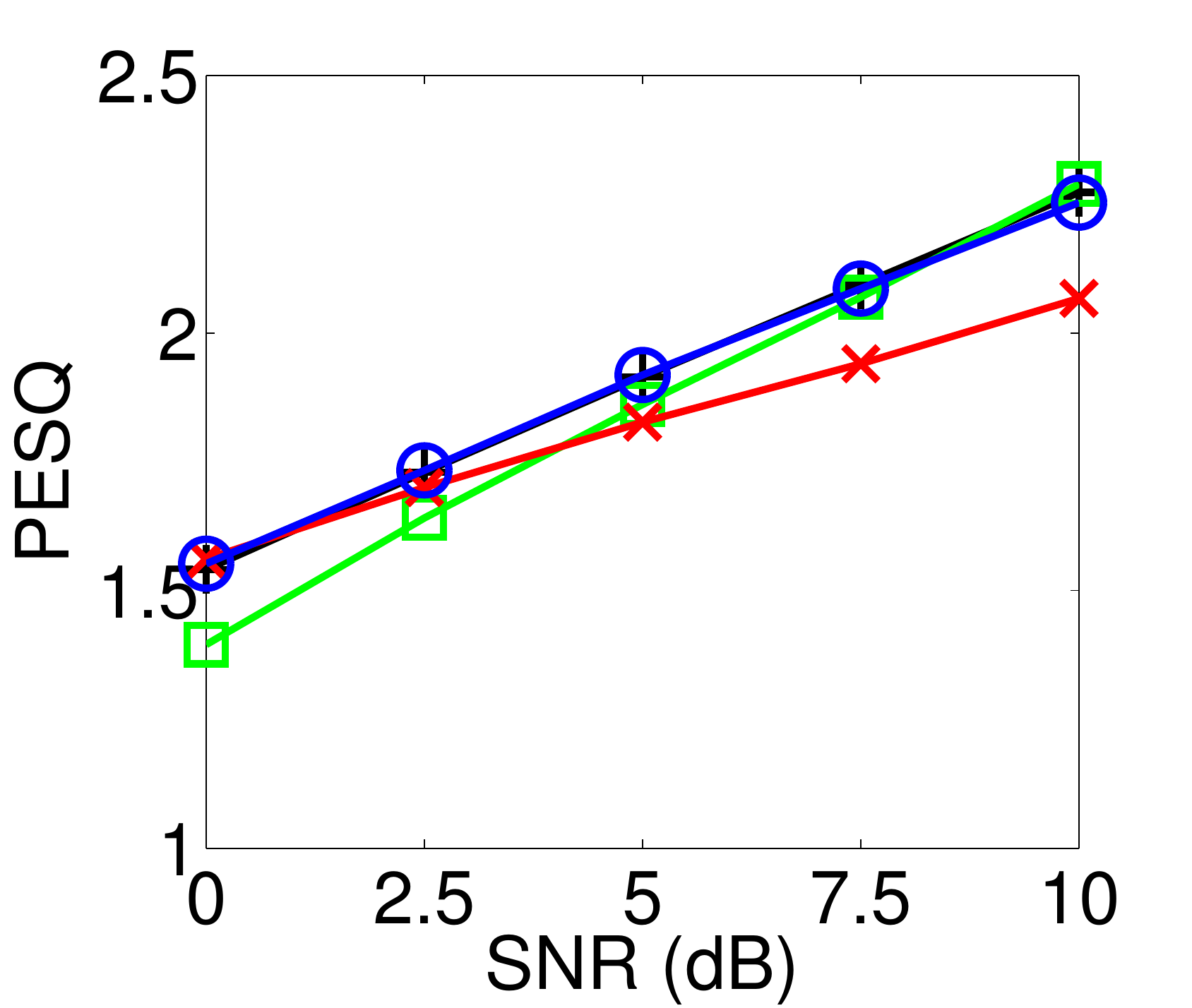}
\caption{Train}
\end{subfigure}
\begin{subfigure}{0.19\textwidth}
\includegraphics[height=2.9cm]{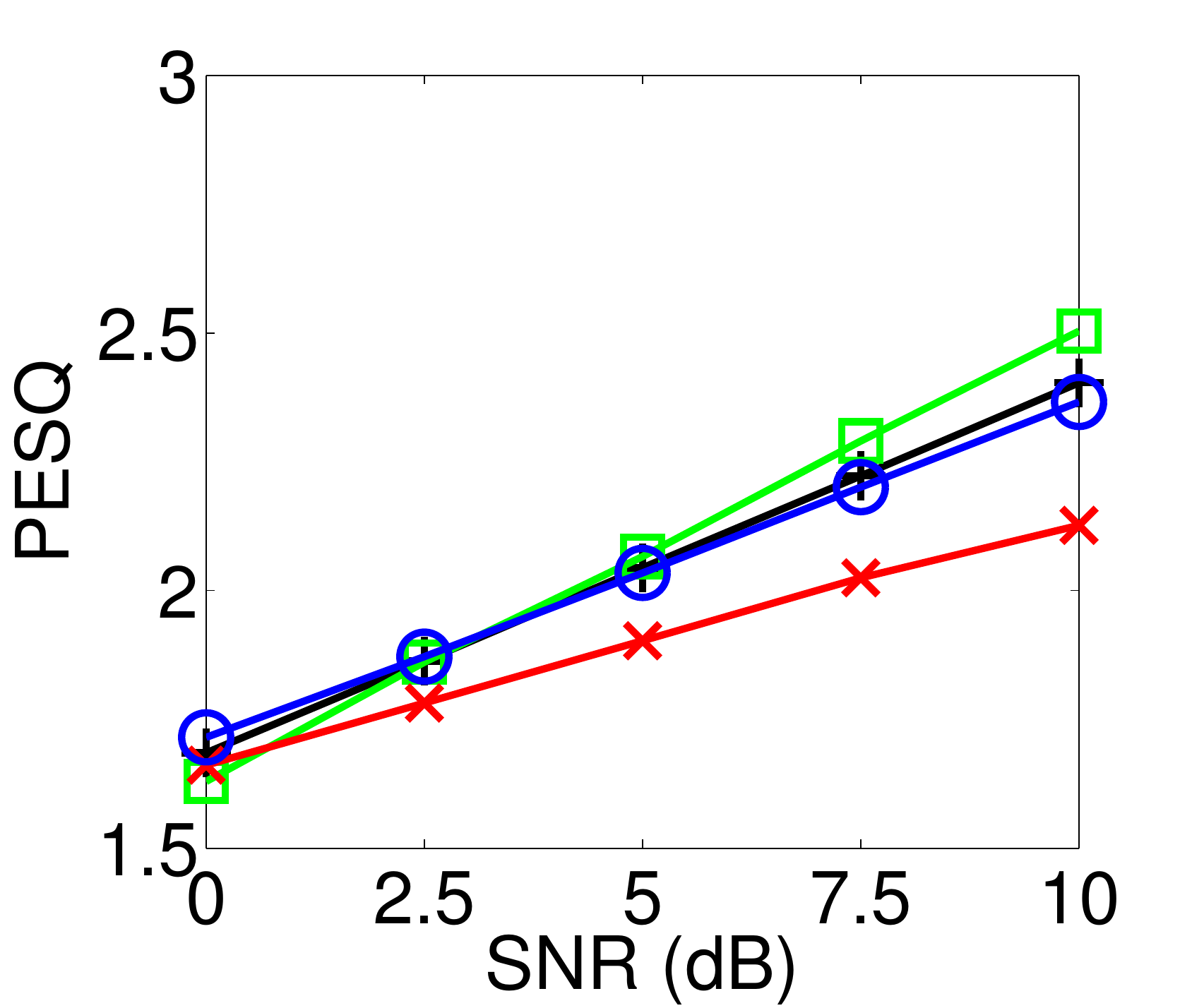}
\caption{Wind}
\end{subfigure}
\begin{subfigure}{0.19\textwidth}
\includegraphics[height=2.9cm]{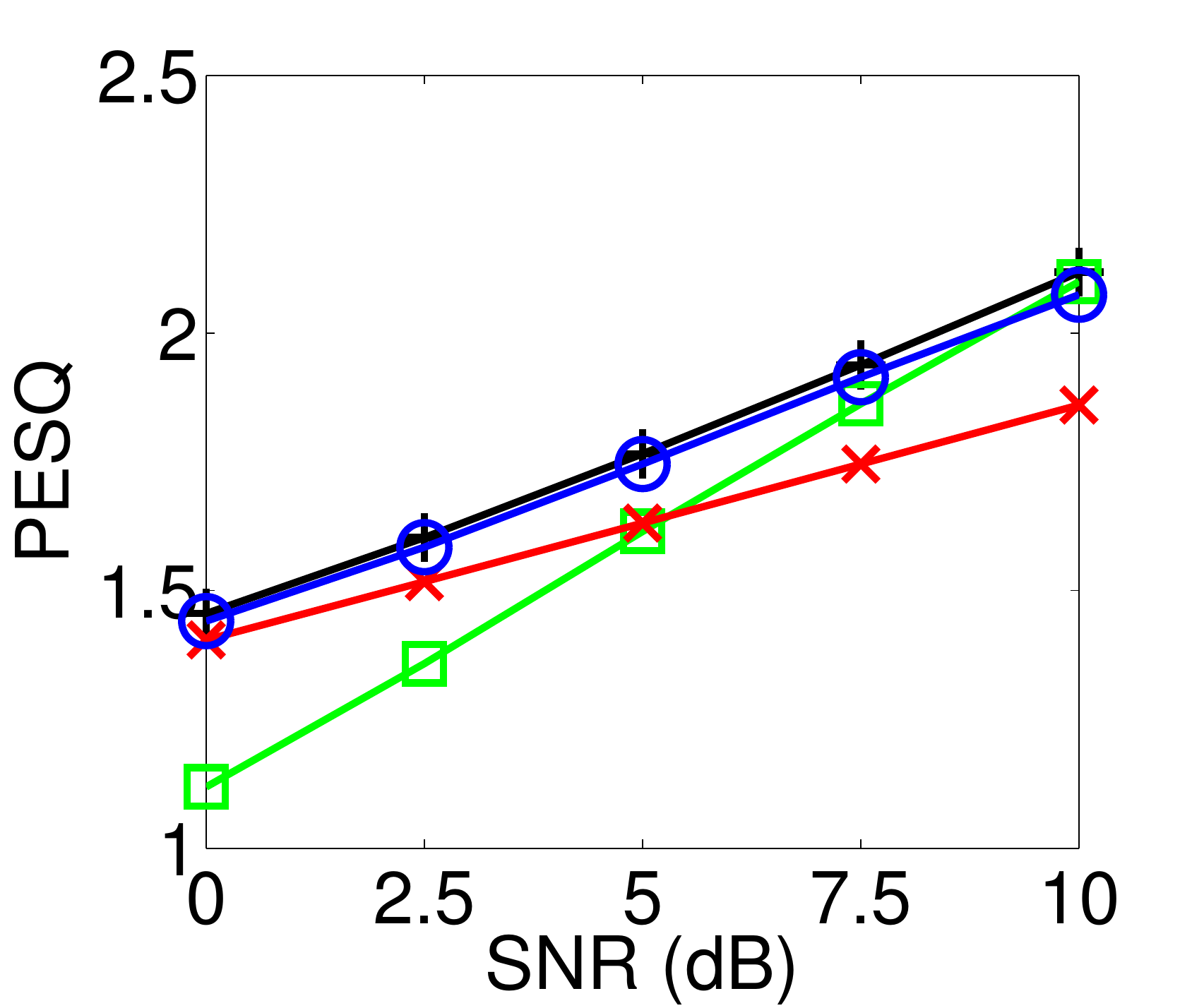}
\caption{Crowd babble}
\end{subfigure}
\vspace{-1mm}
\caption{PESQ on enhanced speech output.}
\label{fig:pesq}
\end{figure*}
        
\begin{figure*}
\centering
\begin{subfigure}{0.25\textwidth}
\includegraphics[width=\textwidth]{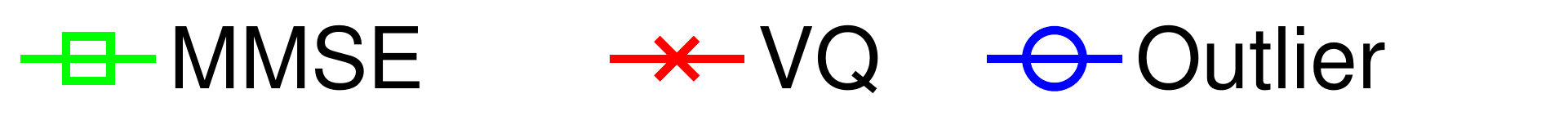}
\end{subfigure} \\
\begin{subfigure}{0.19\textwidth}
\includegraphics[height=2.9cm]{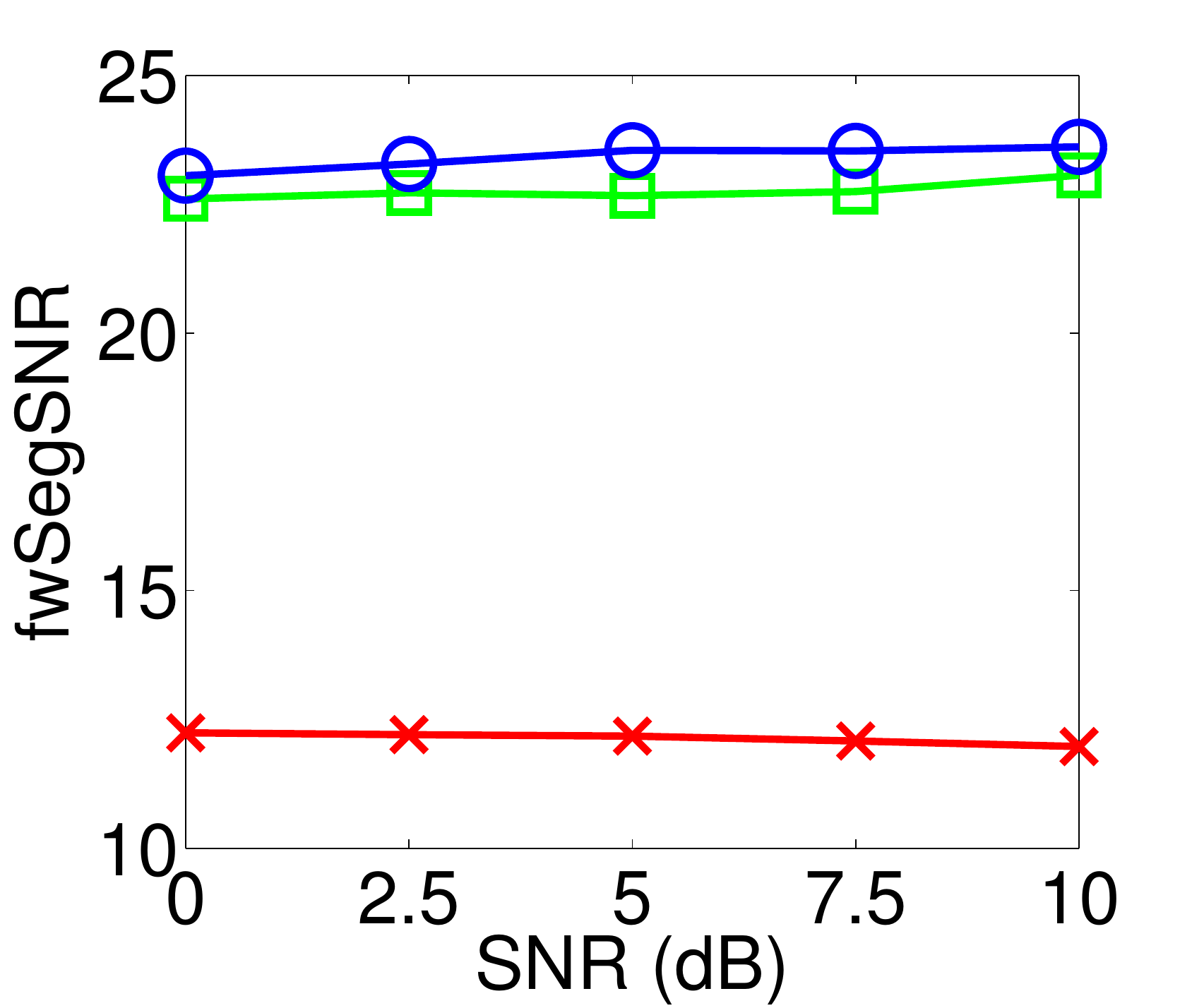}
\caption{Bird}
\end{subfigure}
\begin{subfigure}{0.19\textwidth}
\includegraphics[height=2.9cm]{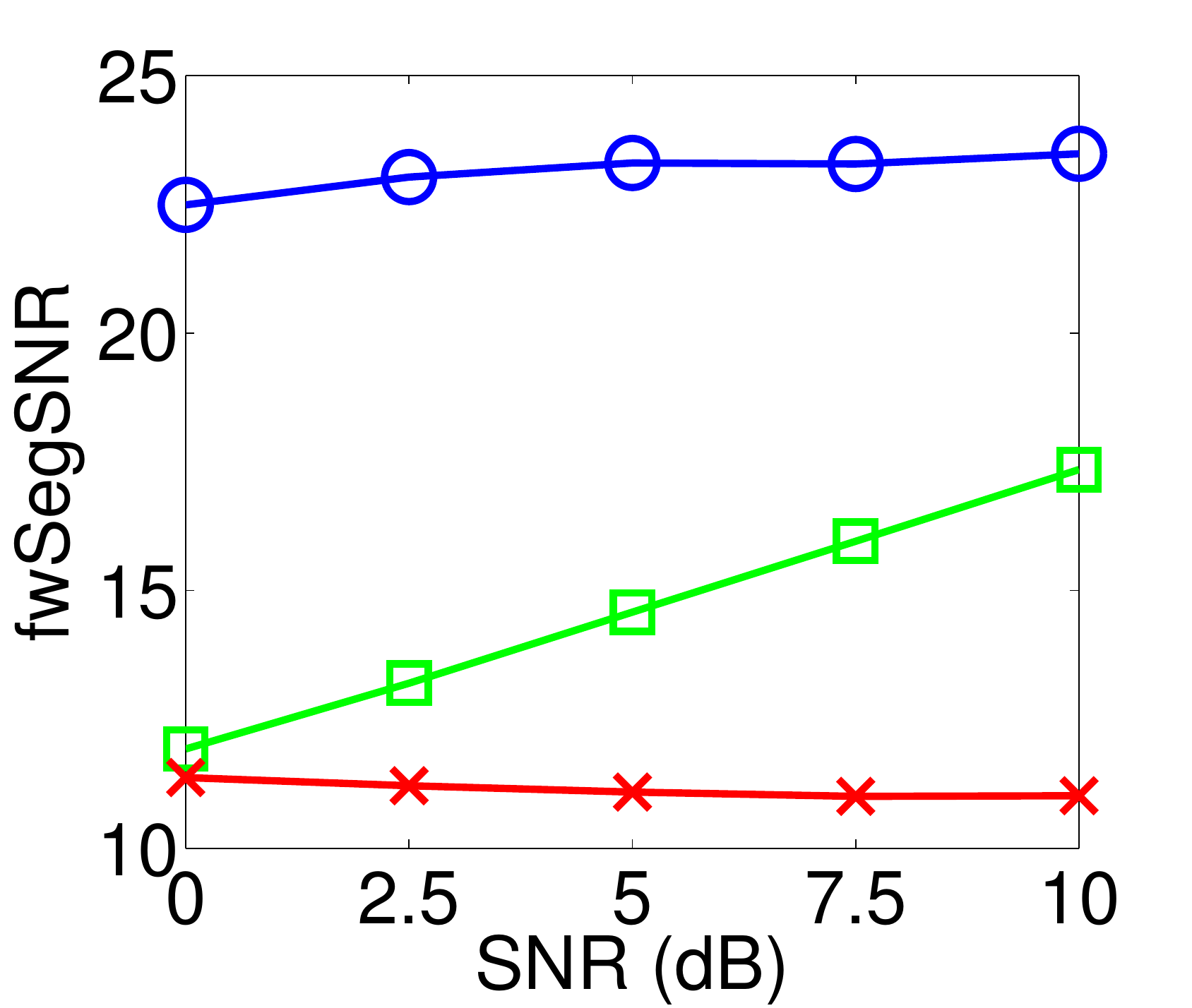}
\caption{Siren}
\end{subfigure}
\begin{subfigure}{0.19\textwidth}
\includegraphics[height=2.9cm]{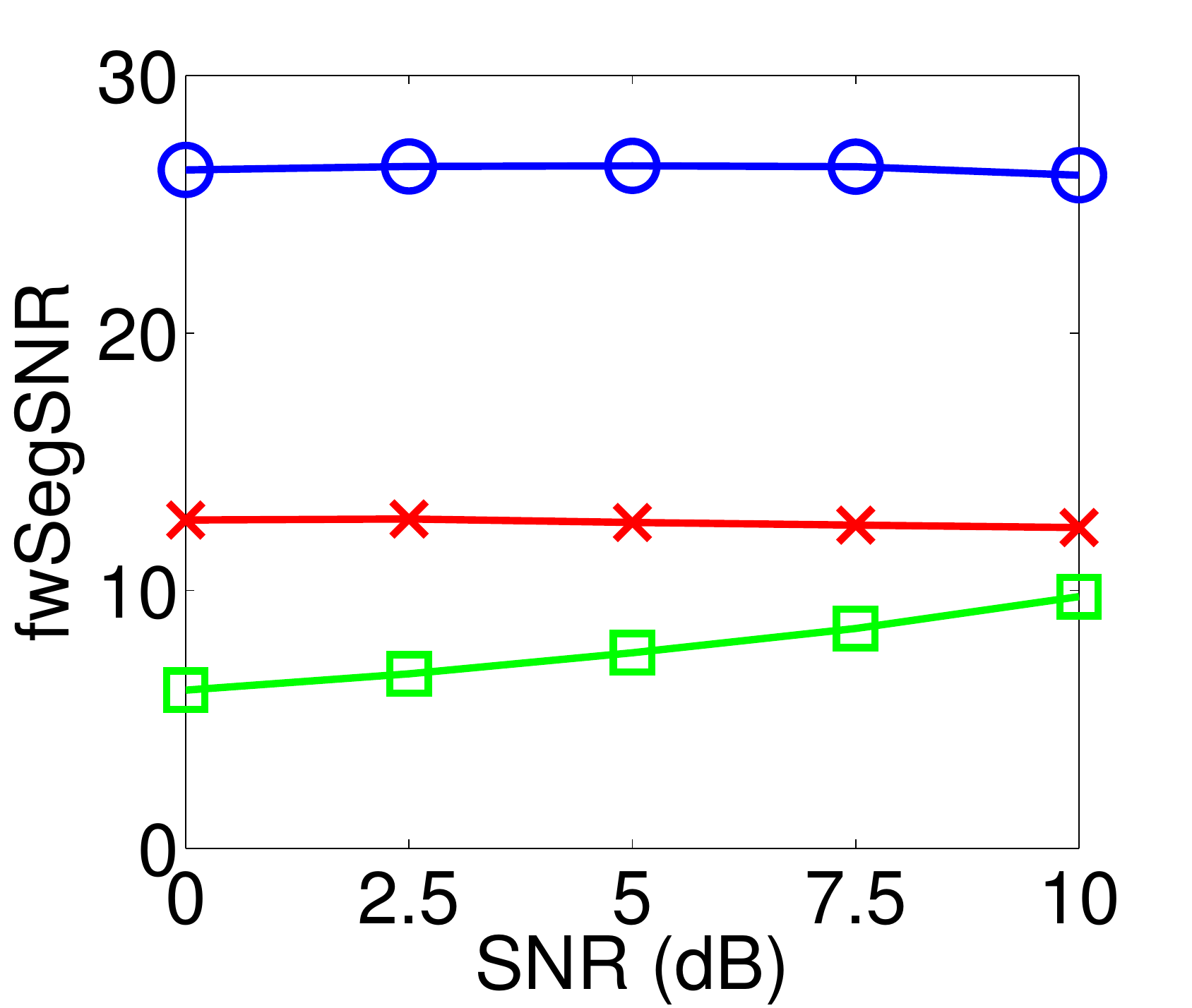}
\caption{Train}
\end{subfigure}
\begin{subfigure}{0.19\textwidth}
\includegraphics[height=2.9cm]{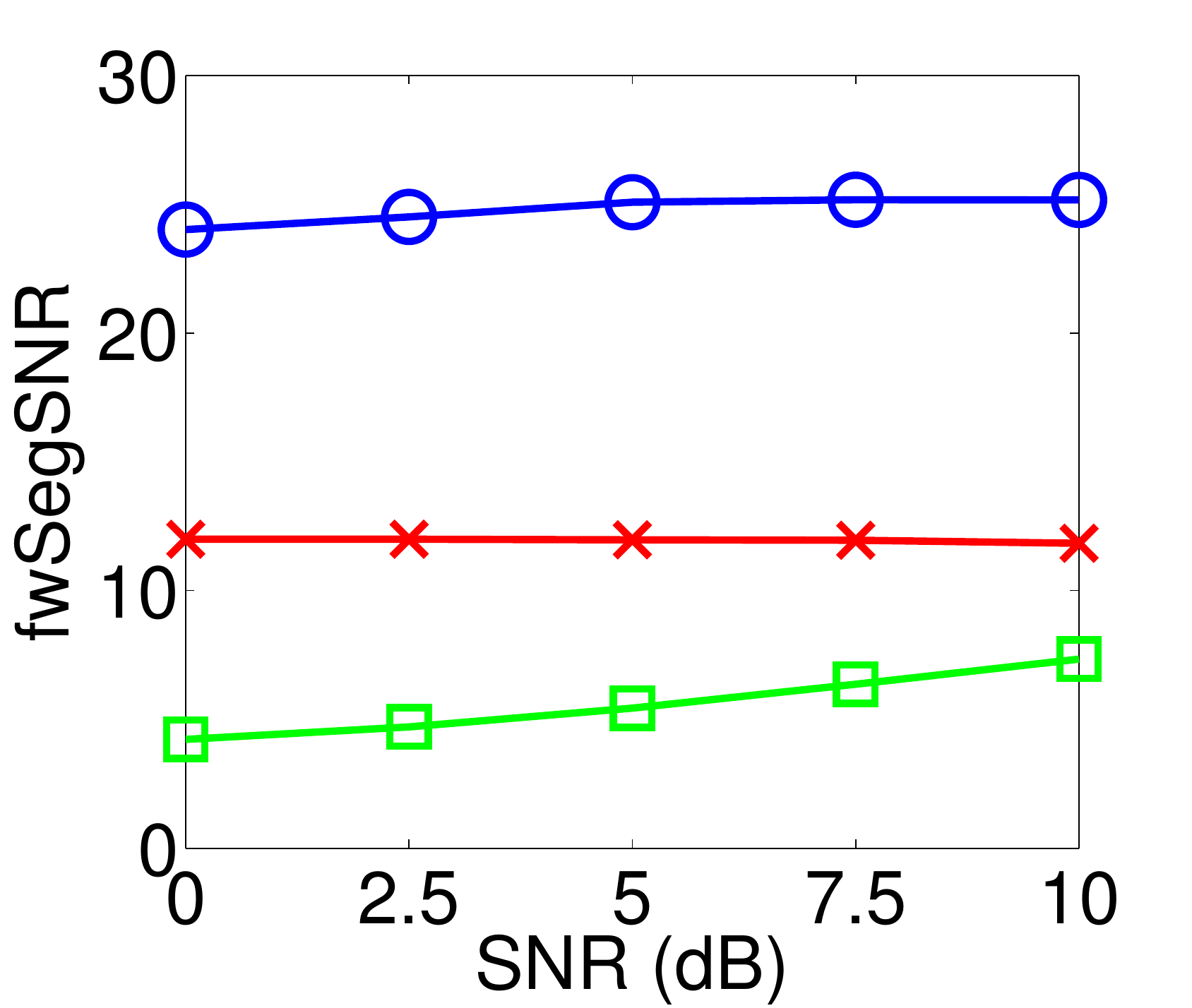}
\caption{Wind}
\end{subfigure}
\begin{subfigure}{0.19\textwidth}
\includegraphics[height=2.9cm]{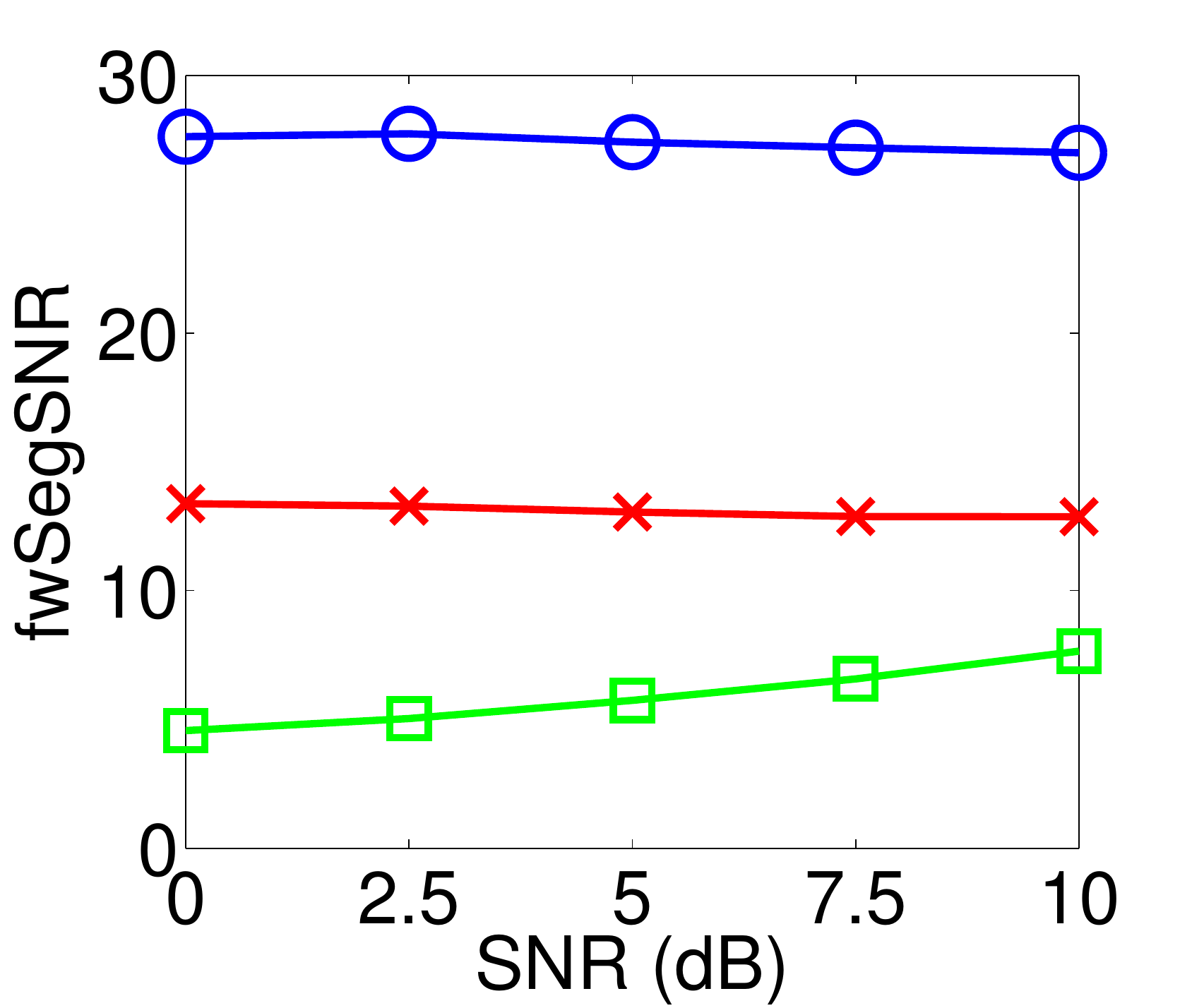}
\caption{Crowd babble}
\end{subfigure}
\vspace{-1mm}
\caption{fwSegSNR on underlying clean speech output.}
\label{fig:energy}
\vspace{-1mm}
\end{figure*}

Two methods, both of which assume no knowledge of the noise, are used for comparison. The first algorithm is based on Ellis's method~\cite{ellis06} where a VQ representation of the speech is used to quantize the noisy speech. To independently evaluate the effect of the outlier framework, only Eq.~\eqref{eq:mask} is replaced with $F_H(i) = \hat F_X(i)/F_Y(i)$. In other words, the mask is computed based on the best selected dictionary entry. The second algorithm is the MMSE noise estimation algorithm by Gerkmann used with an a-priori SNR estimated Wiener filter~\cite{gerkmann12}. Unlike our method, the MMSE algorithm relies on no prior knowledge of speech in general. However, given that the MMSE algorithm has recently shown to be one of the more effective methods for single channel speech enhancement~\cite{taghia11}, we compare it here to highlight some of the more challenging situations a speech enhancement system can encounter. A smoothing factor of $\alpha$ = 0.98 was used for the a-priori SNR estimation. Our outlier method was used with a threshold of $c = 0.0001$.

To train the speech dictionary, 10,000 patches were sampled from randomly selected sentences in the training section of the TIMIT speech database~\cite{timit}. These sentences consisted of both male and female speakers with various accents. 60 filter bands ($N'$) were used to reduce the frequency dimension of the patches, and various patch lengths ($L$ = 1, 2, 4, 8) were evaluated. A similar search was done for other parameters such as the dictionary size ($K$ = 100, 200, 400, 800, 1600) and the analysis window length (5, 10, 15, 20, 25 ms) used to compute the STFT. The parameters that maximized our results ($L$ = 2, $K$ = 800, window length = 10 ms) are shown here. The effect of these parameters are analyzed in~\cite{cho13}.

In the literature, a longer analysis window is often used~\cite{duan12, sigg12} such that the harmonics are better defined. This makes the entries more incoherent with the noise and can lead to better separation when mixed with wideband noise. However, by using a shorter window the formant structure is emphasized, and this allows a smaller number of dictionary entries to capture the multi-speaker training data.

To generate the noisy speech, 5 different noise sources (bird, siren, train, wind and crowd babble) available online~\cite{grsites}, were mixed with sentences from the TIMIT database. 10 sentences from 10 different speakers (5 male, 5 female), not included in the training data, were used to generate these noisy sentences. The average scores on the test set are provided for each of the experiments discussed below.

Fig.~\ref{fig:pesq} shows the Perceptual Evaluation of Speech Quality (PESQ) for the enhanced output. The benefit of using a model based on speech is clear for bird and siren noises. When the spectral shape of the noise is incoherent with speech, a great amount of noise reduction is achievable. However, when the noise is strongly mixed with speech, enhancement is much more challenging. This is shown by the PESQ gains for train, wind and crowd babble noise. The gain is minimal or even a loss in perceptual quality is often observed.

The benefit of the outlier method is highlighted in Fig.~\ref{fig:energy} where the distortion of the underlying speech is measured. The mask in Eq.~\eqref{eq:mask} is initially computed from a noisy speech input. Then, only the underlying clean speech is fed through this computed mask to measure how the speech is affected by the algorithm. The frequency-weighted segmental SNR (fwSegSNR)~\cite{tribolet78} is another measure known to highly correlate with subjective mean opinion scores (MOS)~\cite{loizou07}. This is used to measure the quality of the clean speech output. Fig.~\ref{fig:energy} shows that the outlier method is consistent in preserving the underlying speech regardless of the environment. The framework only reduces the noise when it is possible to do so without distorting the speech.
Being able to limit the amount of speech distortion regardless of the noise encountered is a key benefit of the outlier framework.

\section{Conclusion} \label{sec:conclusion}
A speech enhancement method that is generally applicable to various noisy environments without extensive modification for each environment can be useful. We achieve this by first removing noise components that strongly differ from the trained speech model, and passing the noisy speech when in doubt. Using this outlier framework we are able to greatly reduce noises that are incoherent with speech even if they are non-stationary. Moreover, in environments where separation of the mixed sources is difficult,  speech distortion is minimized. This allows one to use this system as a pre-processing step for various speech processing/recognition applications without much worry of distorting the underlying speech.

\newpage
\bibliographystyle{IEEEbib}
\bibliography{outlier}

\end{document}